\documentclass[aps,prb,twocolumn,superscriptaddress]{revtex4}
\usepackage{dcolumn} 
\usepackage{graphicx}
\usepackage{bm}      
\usepackage{xcolor}
\usepackage{amsmath,amssymb}
\usepackage{hyperref}
\usepackage{multirow}
\usepackage{epstopdf}
\usepackage{latexsym}
\usepackage{subfigure}
\usepackage{wasysym} 
\usepackage{times}
\usepackage{gensymb}
\usepackage{textcomp}

\usepackage[normalem]{ulem}

\begin{document}
\title{Magnetic phase diagram of $A_{2}$[FeCl$_{5}$(H$_{2}$O)] ($A$ = K, Rb, NH$_{4}$)}
\author{Minseong Lee}
\email{minseong.lee10k@gmail.com}
\affiliation{National High Magnetic Field Laboratory, Los Alamos National Laboratory, Los Alamos, NM 87545, USA}
\date{\today}

\begin{abstract}
Erythrosiderites with the formula $A_{2}$Fe$X_{5}\cdot$H$_{2}$O, where $A$ = Rb, K, and (NH$_{4}$) and $X$ = Cl and Br are intriguing systems that possess various magnetic and electric phases, as well as multiferroic phases in which magnetism and ferroelectricy are coupled. In this report, we study the magnetic phase diagram of erythrosiderites as a function of superexchange interactions and magnetic anisotropies. To this end, we perform classical Monte Carlo simulations on magnetic Hamiltonians that contain five different superexchange interactions with single-ion anisotropies. Our phase diagram contains all magnetic ground states that have been experimentally observed in these materials. We argue that the ground states can be explained by varying the ratio of $\frac{J_{4}}{J_{2}}$. For $\frac{J_{4}}{J_{2}} > 0.95$ a cycloidal spins structure is stabilized as observed in (NH$_{4}$)$_{2}$FeCl$_{5}\cdot$H$_{2}$O and otherwise a collinear spin structure is stabilized as observed in (K,Rb)$_{2}$FeCl$_{5}\cdot$H$_{2}$O. We also show that the difference in the single-ion anisotropy along $a$- and $c$- axes is essential to stabilize the intermediate state observed in (NH)$_{2}$FeCl$_{5}\cdot$H$_{2}$O.
\end{abstract}
\maketitle
\section{INTRODUCTION}
Understanding how different physical degrees of freedom couple to each other to induce intricate ground states and how these ground states respond to external stimuli is of fundamental importance to creating new multifunctional materials \cite{Intro1}. For example, magnetoelectric multiferroics are compounds in which magnetic and electric orders couple such that magnetic field can control ferroelectricity and electric field can control magnetism \cite{Intro2}. The past two decades have witnessed an explosion in the number of known magnetoelectric and multiferroic materials and concomitant advances in  understanding their magnetoelectric coupling mechanisms \cite{Intro3}. Most microscopic mechanisms for magnetoelectric coupling are classified into one of three categories \cite{Intro4}: magnetostriction, inverse Dzyaloshinskii-Moriya (DM) \cite{inverseDM1,inverseDM2}, and p-d hybridization \cite{pdhybridization1, pdhybridization2}. Recent discovery of multiferroic properties in compounds with transition metal ions and organic ligands has expanded the arena of candidate materials significantly. In particular, the relatively soft lattice of such compounds due to organic linkers between magnetic polyhedra allows a giant susceptibility to external stimuli that is unattainable in inorganic compounds. Modest field and pressure can readily induce unusual phases. Notably, organic ligands allow us to finely tailor magnetic exchange interaction via ligand engineering. In all cases, understanding the mechanism in new multiferroic materials begins with identifying the magnetic exchange interactions and their role in determining the ground state spin structure. 

\begin{figure*}[tbp]
	\linespread{1}
	\par
	\begin{center}
		\includegraphics[width=0.9\textwidth]{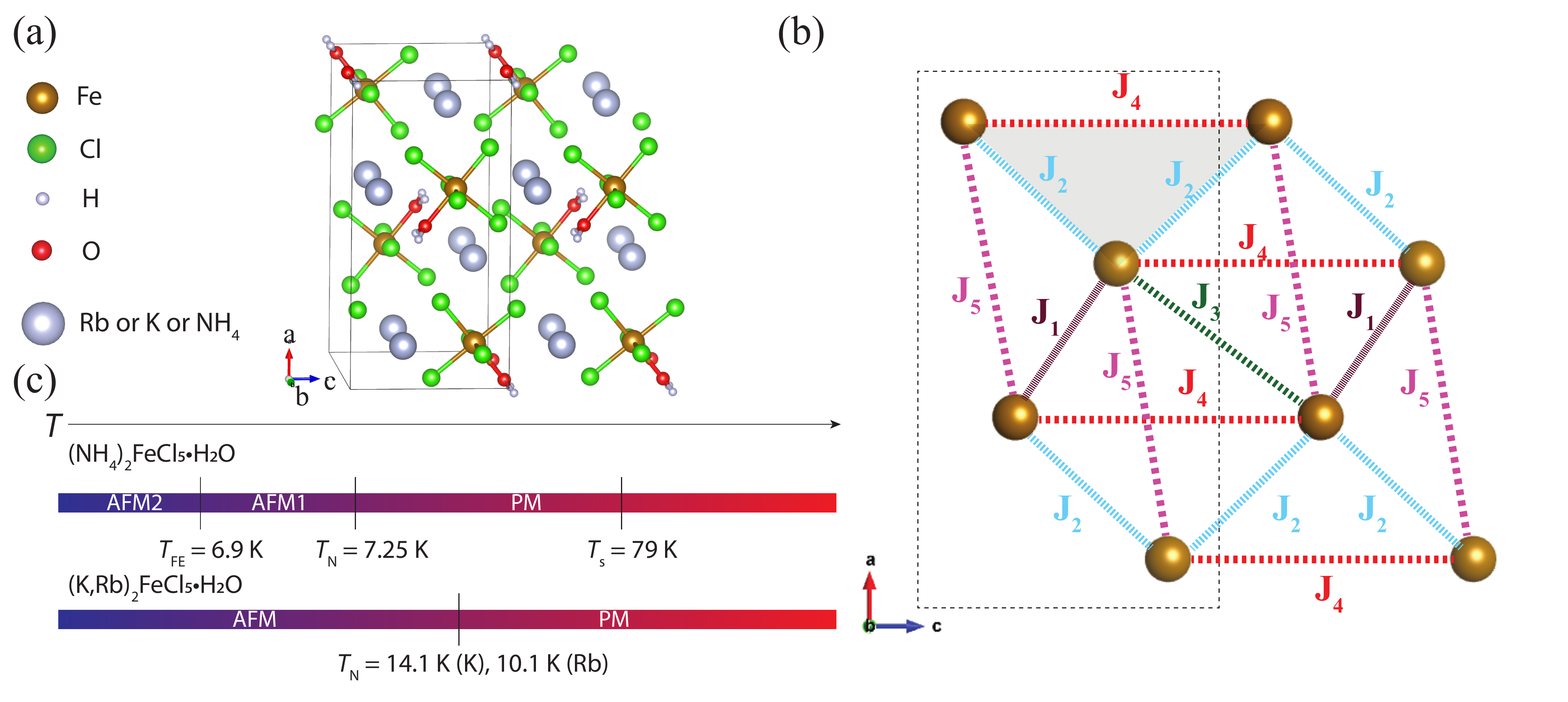}
	\end{center}
	\par
	\caption{\label{structure_exchange2} (color online) (a) Crystal structure of $A_{2}$FeCl$_{5}\cdot$H$_{2}$O. (b) The five superexchange interaction paths used in this study. We only show Fe ions for clarity. (c) The sequence of thermal phase transitions observed in (NH$_{4}$)$_{2}$FeCl$_{5}\cdot$H$_{2}$O and (K,Rb)$_{2}$FeCl$_{5}\cdot$H$_{2}$O.}
\end{figure*}

Erythrosiderites with the formula $A_{2}$Fe$X_{5}\cdot$H$_{2}$O, where $A$ = Rb, K, and (NH$_{4}$) and $X$ = Cl and Br are intriguing systems that possess complex magnetic and electric phase diagrams including multiferroic and magnetoelectric properties. When $A$ is an alkali metal such as Cs, Rb, and K with $X$ = Cl, a single magnetic phase transition from a paramagnetic (PM) to an antiferromagnetic (AFM) phase is observed at zero magnetic field as schematically shown in \autoref{structure_exchange2} (c). As the radius of the alkali ion increases from K to Rb to Cs, the N\'eel temperature $T_{N}$ decreases from 14.1 K to 10.1 K to 6.5 K \cite{alkali1,alkali2,alkali3}. According to neutron diffraction studies on these alkali compounds, four spins in the unit cell, defined as dashed line box in \autoref{structure_exchange2} (a) form antiferromagnetic order with spins along the $a$ axis \cite{alkali_neutron1}. An inelastic neutron scattering study suggests that the magnetic Hamiltonian includes at least five exchange interactions as denoted in \autoref{structure_exchange2} (b) with small magnetic anisotropies \cite{inelastic_alkali}. In applied magnetic fields, a spin-flop phase transition occurs between 1 T and 3 T and no additional magnetic phase appears up to 16 T \cite{alkali4,alkali5,alkali6}. Although the collinear AFM phase is not multiferroic at zero magnetic field, all three compounds exhibit a linear magnetoelectric effect as electric polarization is induced in applied magnetic fields that persists in the canted spin phase \cite{alkali6}. Such a linear magnetoelectric effect requires the system to break time-reversal as well as spatial inversion symmetry.

When $A$ is a NH$_{4}$ molecule and $X$ = Cl, even more interesting magnetic and electric phases emerge. First, this compound shows a structural phase transition around $T_{s}$ = 79 K corresponding to the order-disorder transition of NH$_{4}$ freezing its rotational degree of freedom \cite{Ruby_79K_1,Ruby_79K_2}. An incommensurate sinusoidal collinear order of the $a$ component of the Fe spin develops at $T_{N}$ = 7.25 K with propagation vector ${\bf k} \sim $ (0, 0, 0.23) \cite{Ruby_spin_lattice}. At $T_{\text{FE}}$ = 6.9 K, another magnetic phase transition from the sinusoidal phase to an incommensurate cycloidal spin structure is observed, where the spiral plane of the $ac$-plane propagates along the $c$-axis with $k_{c} \sim $ 0.23 \cite{Ruby_79K_1,Ruby_spin_lattice}. Ferroelectricity along the $a$-axis emerges concurrently with the spin cycloidal phase, hence (NH$_{4}$)$_{2}$FeCl$_{5}\cdot$H$_{2}$O has a multiferroic ground state. Interestingly, this series of phase transitions closely resembles the sequence of phase transitions in the well-known orthorhombic rare-earth manganites\cite{TbMnO3_1}. Multiferroic behavior in these compounds is attributed to the inverse-DM mechanism because the direction of the spontaneous electric polarization is perpendicular to both the spin chirality ${\bf S}_{i}\times {\bf S}_{j}$ and spin propagation directions \cite{Ruby_79K_1}. However, in contrast to the orthorhombic rare-earth manganites, in erythrosiderites the DM interaction between nearest-neighbor spins is not allowed by symmetry and thus the spin cycloid phase is controlled only by a combination of isotropic exchange interactions \cite{xiaojian_1}. 
In a recent neutron diffraction measurement to resolve the spin structure, the distorted spin cycloidal spin structure is described by not only the odd harmonics (2$n$-1)$k_{c}$ but also the rather strong second harmonics ($2n$)$k_{c}$. The odd harmonics suggests the magnetic anisotropic in $ac$-plane \cite{randybook} and the even harmonic implies a strong spin-lattice coupling \cite{Ruby_spin_lattice}.

One of fascinating properties of (NH$_{4}$)$_{2}$FeCl$_{5}\cdot$H$_{2}$O is the control of the direction of electric polarization using the magnetic fields. As the magnetic field along $a$ or $c$ axis increases, the electric polarization 
rotates away from $a$ axis by up to 10 degrees and then flips to the $c$ axis above the spin-flop phase transition field. \cite{Ruby_79K_2} The inverse-DM mechanism for multiferroicity can no longer account for the direction of the electric polarization in the high field phase. It has been suggested that the $p$-$d$ hybridization mechanism explains the electric polarization properly in high magnetic fields above 5 T \cite{mechanismchange1}. This makes (NH$_{4}$)$_{2}$FeCl$_{5}\cdot$H$_{2}$O a unique system that switches the microscopic mechanism of multiferroicity under magnetic fields. The origin of the mechanism change remains unknown but the magnetic exchange interaction is robust regardless of the multiferroic mechanism \cite{xiaojian_1}. 

More recently, thermodynamic properties of the solid solution of (NH$_{4}$)$_{2-x}$K$_{x}$FeCl$_{5}\cdot$H$_{2}$O has been reported \cite{solidsolution1}. The increases in K content raises $T_{N}$ monotonically and transforms the cycloidal spin state for $x < 0.06$ into the collinear spin state for $x > 0.15$. Ferroelectricity induced by the noncollinear spin structure is accordingly suppressed drastically by postassium. This implies that (NH$_{4}$)$_{2}$FeCl$_{5}\cdot$H$_{2}$O could be the most frustrated system among the series of erythrosiderites compounds.

Despite all these unusual phenomena, an understanding of how the magnetic states such as cycloidal, AFM sinusoidal and collinear AFM are determined as a function of $A$ is missing. It has been argued that geometrical frustration on the triangular plaquette formed by $J_{4}$ and $J_{2}$ mainly stabilizes the cycloidal spin structure in (NH$_{4}$)$_{2}$FeCl$_{5}\cdot$H$_{2}$O compound in the absence of DM interactions \cite{Ruby_79K_1}. On the other hand, the reduced $J_{4}$ interaction due to the absence of hydrogen bonds relieves the frustration and stabilizes the collinear AFM ground state (K, Rb)$_{2}$FeCl$_{5}\cdot$H$_{2}$O compounds \cite{inelastic_alkali}. As shown in \autoref{energy_table1}, a slightly higher value of $J_{4}/J_{2}$ ratio of NH4 compounds than that of K and Rb compounds also support the arguments. To prove these arguments, we theoretically investigate the magnetic phase diagram of A$_{2}$FeCl$_{5}\cdot$H$_{2}$O at zero magnetic in this report. We used the minimal Heisenberg spin Hamiltonian with five exchange interactions including single-ion anisotropies. We performed Monte Carlo simulations to obtain the magnetic phase diagram. Our numerical studies shows that although there could be seemingly many intricate combinations of five exchange interactions, a single parameter of the ratio of $\frac{J_{4}}{J_{2}}$ could tune the magnetic ground state from noncollinear spin spiral to collinear spin structure. This result supports the speculation that the triangular plaquette formed by $J_{4}$ and $J_{2}$ shown in \autoref{structure_exchange2} (b) gives rise to the magnetic frustration that stabilizes the cycloidal spin structure in (NH$_{4}$)$_{2}$FeCl$_{5}\cdot$H$_{2}$O compound in the absence of DM interactions \cite{Ruby_79K_1}. In addition, we also found that magnetic anisotropy in the $ac$-plane is required to reproduce collinear spin structure along $a$-axis when the magnetic frustration is relieved for $A$ being alkali ions, and the sinusoidal AFM phase observed in (NH$_{4}$)$_{2}$FeCl$_{5}\cdot$H$_{2}$O.


\section{Method and Model Hamiltonian}

\begin{figure}[tbp]
	\linespread{1}
	\par
	\begin{center}
		\includegraphics[width=0.45\textwidth]{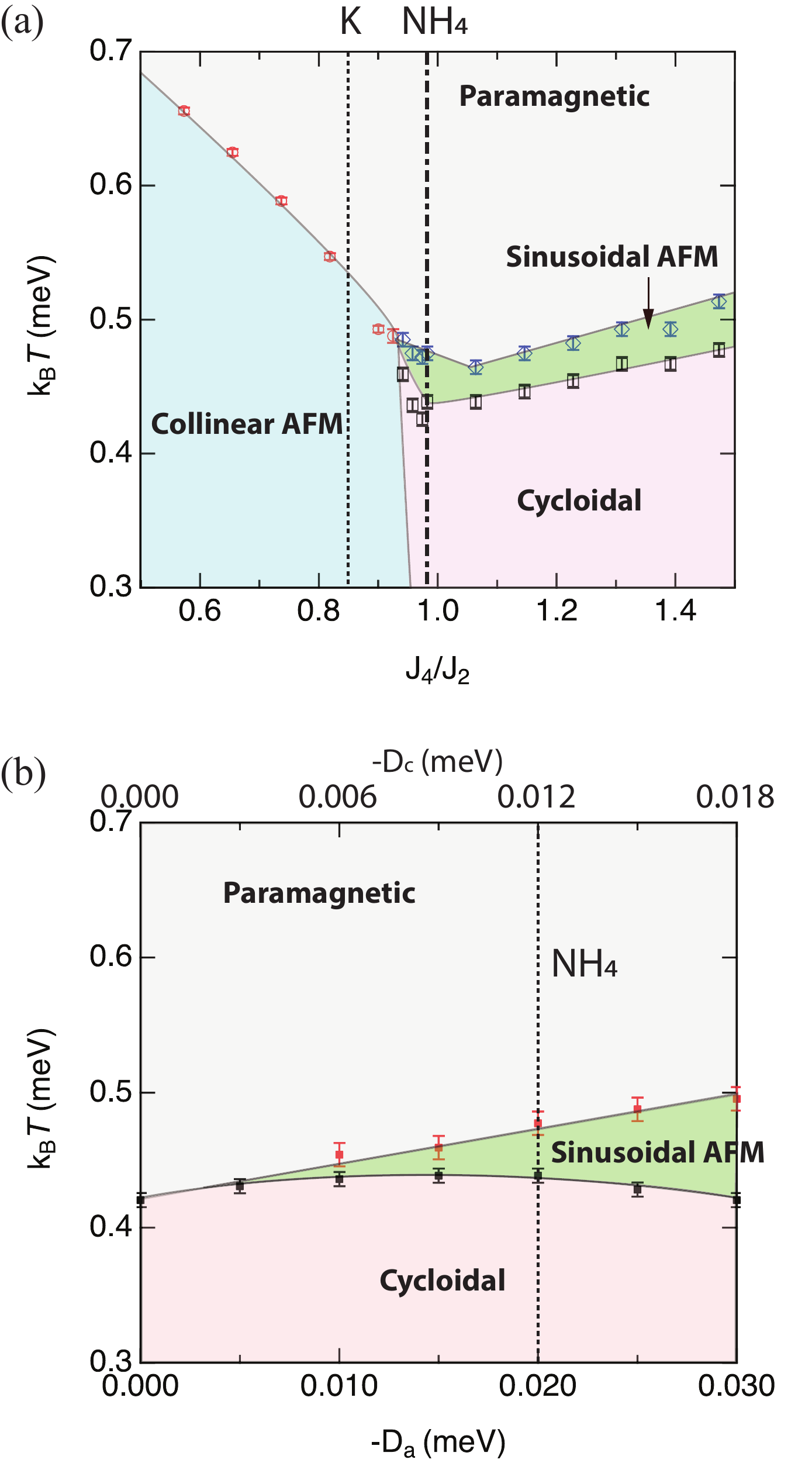}
	\end{center}
	\par
	\caption{\label{PD1} (color online) Magnetic phase diagrams from Monte Carlo simulations at zero magnetic field (a) as a function of the ratio $J_{4}~/J_{2}$ and of the easy axis anisotropy $D_{a}$ and $D_{c}$.
}
\end{figure}

\begin{table*}[]
\caption{\label{energy_table1}Comparison of exchange parameters and single-ion anisotropies of $A_{2}$Fe$X_{5}\cdot$H$_{2}$O, where $A$ = Rb, K, and (NH$_{4}$) and $X$ = Cl and Br. Unit in meV. MCS = Monte Carlo Simulation, INS = Inelastic Neutron Scattering experiments.}
\begin{tabular}{llllllllllll}
\hline\hline
Composition & J$_{1}$ & J$_{2}$ & J$_{3}$ & J$_{4}$ & J$_{5}$ & D$_{a}$ & D$_{b}$ & D$_{c}$ & J$_{4}$/J$_{2}$ & Method                       & Reference \\
\hline\\
(NH$_{4}$)$_{2}$FeCl$_{5}\cdot$H$_{2}$O  & 0.1604   & 0.0611   &  0.0301  &  0.06  &  0.03191  &     -0.02& 0    & -0.012    &  0.98     & MCS      & This work \\
            & 0.1710(1)   & 0.05871(6)    &0.02985(4)    & 0.0535   & 0.0461   & 0    & 0.01043    &   0  &    0.91   & INS & \cite{xiaojian_1}  \\
            & 0.1801(2)   & 0.05313(14)   & 0.03198(15)   &  0.05535(11)  &  0.04001(8)  &   0  &0.01543(6)     & 0   &1.04       & INS &  \cite{xiaojian_1}  \\
            & 0.217   & 0.1341   & 0.0224   & 0.1082   & 0.06168   &    -0.01 &  0   & -0.006    &  0.81     & {\it ab initio} & \cite{amanda1}     \\
K$_{2}$FeCl$_{5}\cdot$H$_{2}$O           & 0.1604   &  0.0611  & 0.0301   & 0.0513   &  0.03191  &  -0.02   & 0    &-0.012     &    0.84   & MCS & This work \\
            &  0.113(3)  & 0.037(1)   & 0.032(3)   &  0.023(2)  & 0.025(2)   & -0.0146    & 0.0072    &  -0.0072   &    0.62   & INS & \cite{inelastic_alkali}       \\
            & 0.142   & 0.086   & 0.084   & 0.071   & 0.052   &  -   &  -   &  -   &   0.83    & {\it ab initio} & \cite{inelastic_alkali}       \\
Rb$_{2}$FeCBr$_{5}\cdot$H$_{2}$O  & 0.131   & 0.152   & 0.154   & 0.138   & 0.063  &  -   &  -   &  -   &  0.91 & {\it ab initio} & \cite{inelastic_alkali}      \\
            \hline\hline
\end{tabular}%
\end{table*}

We performed conventional single-flip Metropolis Monte Carlo simulations to calculate thermodynamic properties of the Hamiltonian Eq.~(\ref{hamiltonian1}). Each temperature involves 2$\times$ 10$^{6}$ Monte Carlo steps, half of which were discarded for thermalization. The results are obtained for 28 $\times$ 10 $\times$ 28 unit cells with periodic boundary condition and each unit cell contains four Fe atoms as described in \autoref{structure_exchange2}. We also performed the same calculations for different sizes and we confirm that the lattice size effect is small for the size used herein.



To describe the magnetic properties of {\it A}$_{2}$[FeCl$_{5}\cdot$(H$_{2}$O)], we used a classical 
 Heisenberg Hamiltonian that consists of isotropic superexchange interactions and single-ion anisotropies and the spins are treated as vectors with a fixed length. Since the energy of the high-spin state of Fe$^{3+}$ is a few eV below the first charge excitation state in an octahedral crystal field, the states with different spin are irrelevant in the temperature range we are interested in in this simulation. Thus, we fixed the spin of Fe$^{3+}$ to $S = 5/2$ by assuming no charge fluctuations exist. We also assume that the system has an orthorhombic crystal structure and the spin quantization axes are identical to the crystal lattice axes. Although the experimental crystal structure is monoclinic, the deviation of the monoclinic angle from 90 degrees is only $\sim$0.1 \% \cite{Ruby_79K_1}.

The model Hamiltonian
\begin{equation}
    \mathcal{H} = \sum_{k}J_{k}\sum_{i,j}'{\bf S}_{i}\cdot {\bf S}_{j} + D_{a}\sum_{i}(S_{i}^{a})^{2} + D_{c}\sum_{i}(S_{i}^{c})^{2}
    \label{hamiltonian1},
\end{equation}
includes all five exchange interactions defined in \autoref{structure_exchange2} (b) and the single ion anisotropies that are different along the $a$, $b$ and $c$-axes. Since $b$ is the hard axis, we set the single-ion anisotropy along $b$ to zero \cite{Ruby_79K_2} and let both $D_{a}$ and $D_{c}$ be negative. The definition of each superexchange interaction in terms of the crystal structure can be found in \cite{inelastic_alkali} and \cite{amanda1}.

The specific heat is calculated by
\begin{equation}
    \frac{C_{s}\left(T\right)}{k_{B}} = \frac{\left(\Delta E\right)^{2}}{k_{B}^{2}T^{2}} = \frac{\left<E^{2}\right> - \left<E\right>^{2}}{k_{B}^{2}T^{2}},
    \label{specificheat}
\end{equation}
where the angled bracket is the thermal average. The order parameter of the collinear AFM parameter is defined by the staggered moments of the four spins in the unit cell as follows.
\begin{equation}
    \left[l_{a}, l_{b}, l_{c}\right] = \left<{\bf S}_{1} - {\bf S}_{2} + {\bf S}_{3} - {\bf S}_{4}\right>/4S.
    \label{staggered}
\end{equation}
To determine the order parameter of the cycloidal structure, we calculated the spin helicity defined as follows.
\begin{eqnarray}
    \nonumber \left[h_{a}^{i},h_{b}^{i},h_{c}^{i}\right] = \frac{\left<{\bf S}_{i}\times {\bf S}_{i+1}\right>}{S^{2}},
    \label{helicity}
\end{eqnarray}
where $i$ is the direction of the propagation wavevector. 
It is known for erythrosiderites that the noncollinear spin structure induces spontaneous electric polarization via the inverse DM interaction. Thus, the plane of the spin rotation and the propagation direction is valuable information to determine the direction of the polarization. The spin-spin correlation function is calculated by
\begin{equation}
    S_{\alpha}\left({\bf q}\right) = \frac{1}{N^{2}}\sum_{i,j}\left<S_{\alpha,i}S_{\alpha,j}\right>e^{i{\bf q}\cdot\left({\bf r}_{i}-{\bf r}_{j}\right)},
\end{equation}
where $\alpha = a,b,c$. The spin-spin correlation function not only reveals the wavelength of each phase but also distinguishes the sinusoidal AFM and cycloidal spin phases. 

All superexchange interactions defined in \autoref{structure_exchange2} (b) are antiferromagnetic. This has been confirmed not only from theoretical studies of {\it ab initio} calculations \cite{amanda1} but also the experimental studies of inelastic neutron scatterings \cite{inelastic_alkali,xiaojian_1}. In this paper, we take the exchange parameters set obtained by fitting the inelastic neutron scattering simulations to the experimental data \footnote{private communications with Randy S. Fishman} tabulated in \autoref{energy_table1} -- except for $J_{4}$. We varied $J_{4}$ from 0.03 meV to 0.12 meV as a tuning parameter of the degrees of frustration. Although a small magnon gap observed in inelastic neutron scattering makes it difficult to identify the difference between $D_{a}$ and $D_{c}$, the isothermal magnetization measurement clearly shows that $\left|D_{a}\right| > \left|D_{c}\right| > \left|D_{b}\right|$ order \cite{Ruby_79K_2}. We set $D_{a}$ = -0.02 meV, $D_{b}$ = 0 meV and $D_{c}$ = -0.012 meV in our calculations. However, as we shall see later, the results do not change qualitatively as long as the order $\left|D_{a}\right| > \left|D_{c}\right| > \left|D_{b}\right|$ is intact. The various exchange parameters found in literature are summarized in \autoref{energy_table1}.

\section{RESULTS}
\autoref{PD1} (a) shows the magnetic phase diagram as a function of temperature and $\frac{J_{4}}{J_{2}}$, obtained from Monte Carlo simulation. \autoref{PD1} (b) is the magnetic phase diagram as a function of temperature $D_{a}$ with the ratio of $D_{c}$ to $D_{a}$ held fixed at 0.6. The magnetic phase boundaries are determined from the peak positions of the specific heat and they are also consistent with the order parameters and spin-spin correlation calculations. As shown in \autoref{PD1} (a), when $J_{4}/J_{2}$ is below 0.95, there is a single phase transition as a function of temperature from the paramagnetic (PM) state to a collinear antiferromagnetic (AFM) phase. The transition temperature monotonically increases with decreasing $J_{4}/J_{2}$. On the other hand, when $J_{4}/J_{2}$ is above 0.95, a two-step magnetic phase transition from the PM state to the collinear sinusoidal AFM and then the cycloidal spin structure is observed as temperature is lowered. \autoref{PD1} (b) shows how this two-step phase transitions evolve as a function of the single-ion anisotropy. The anisotropy in the $ac$ plane stabilizes the intermediate sinusoidal AFM phase. Thus, when the anisotropy in the $ac$ plane is small, the intermediate sinusoidal AFM phase is suppressed and a direct phase transition from the PM to the spin cycloidal phase is also possible.

The $J_{4}/J_{2}$ vs temperature magnetic phase diagram shown in \autoref{PD1} (a) contains all four magnetic phases observed in erythrosiderites: paramagnetic, collinear AFM, sinusoidal AFM and cycloidal. Although there are five different exchange parameters, our calculation shows that $J_{4}/J_{2}$ could be a key parameter to determine the ground state and the sequence of magnetic phase transitions. We marked the $J_{4}/J_{2}$ values that correspond to $A$ = K ($J_{4}/J_{2}$ = 0.84), NH$_{4}$ ($J_{4}/J_{2}$ = 0.98) of $X =$ Cl in \autoref{PD1} (a) and our simulation indeed reproduces the magnetic phases and the phase transition experimentally observed in these compounds: When $A$ = K, a single phase transition is observed and the ground state is collinear AFM phase. On the other hand, when $A$ = NH$_{4}$, the two-step magnetic phase transition appears from PM to sinusoidal to spin cycloidal. Moreover, values of $J_{4}/J_{2}$ found in other experimental and theoretical studies are also well consistent with values of $J_{4}/J_{2}$ found here as shown in \autoref{energy_table1}. In addition, a noticeable difference between $D_{a}$ and $D_{c}$ observed in isothermal measurement for NH$_{4}$ compound \cite{Ruby_79K_2} is essential in stabilizing the sinusoidal AFM state although a small energy gap prevents inelastic neutron scattering experiments from determining the anisotropic energies unambiguously\cite{xiaojian_1}. We now demonstrate how the simulation identified the spin structure of each magnetic phase below.

\begin{figure}[tbp]
	\linespread{1}
	\par
	\begin{center}
		\includegraphics[width=0.45\textwidth]{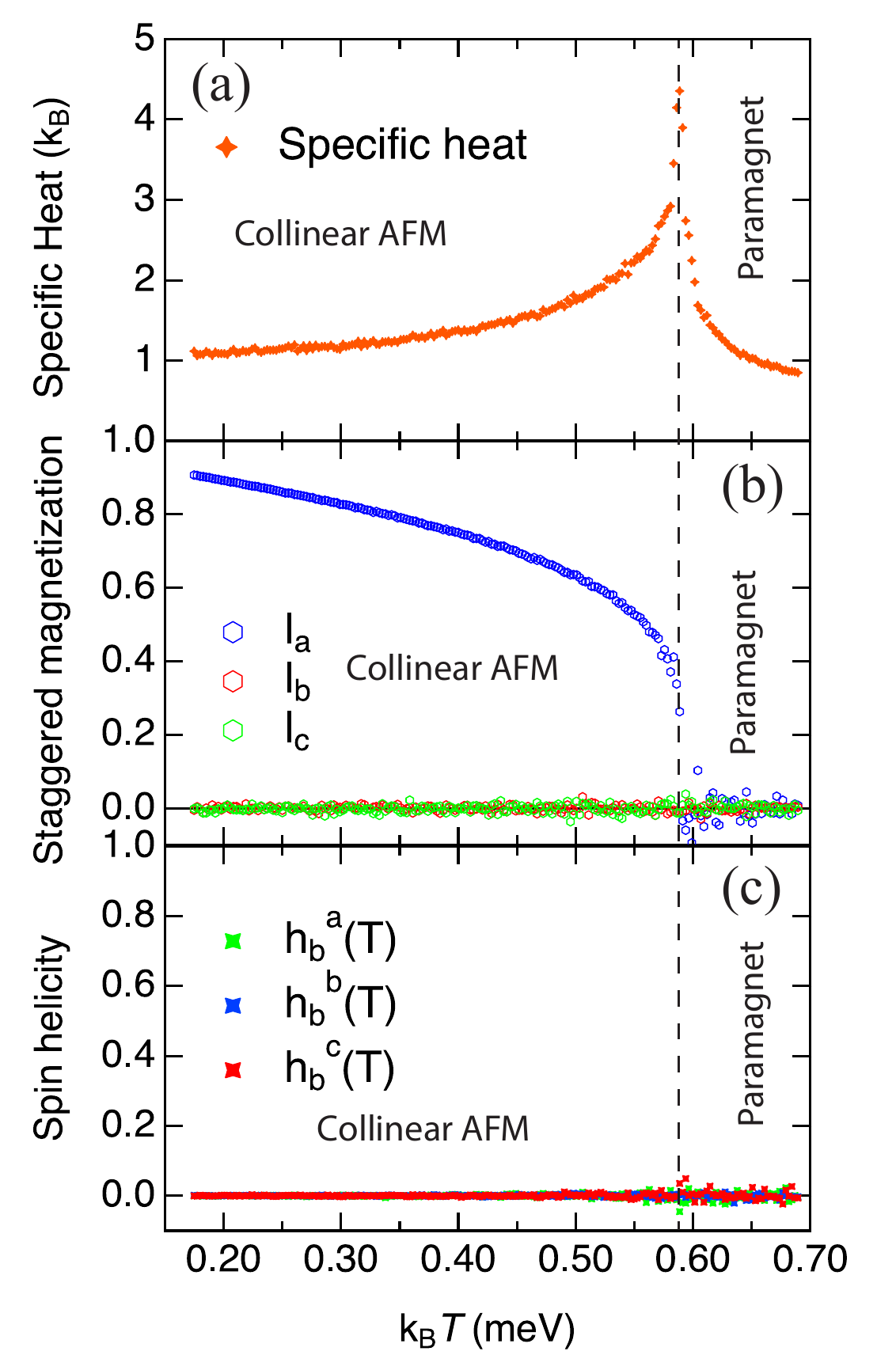}
	\end{center}
	\par
	\caption{\label{structure_exchange3} (color online) Thermodynamic properties of paramagnetic - collinear AFM phase transitions ($J_{4}/J_{2} < $  0.9). Calculated (a) specific heat (Eq.~(\ref{specificheat})) and (b) staggered magnetic moment (Eq.~(\ref{staggered})) and spin helicity (Eq.~(\ref{helicity})) as a function of temperature at zero field at $J_{4}/J_{2}$ = 0.74.}
\end{figure}

\begin{figure}[tbp]
	\linespread{1}
	\par
	\begin{center}
		\includegraphics[width=0.45\textwidth]{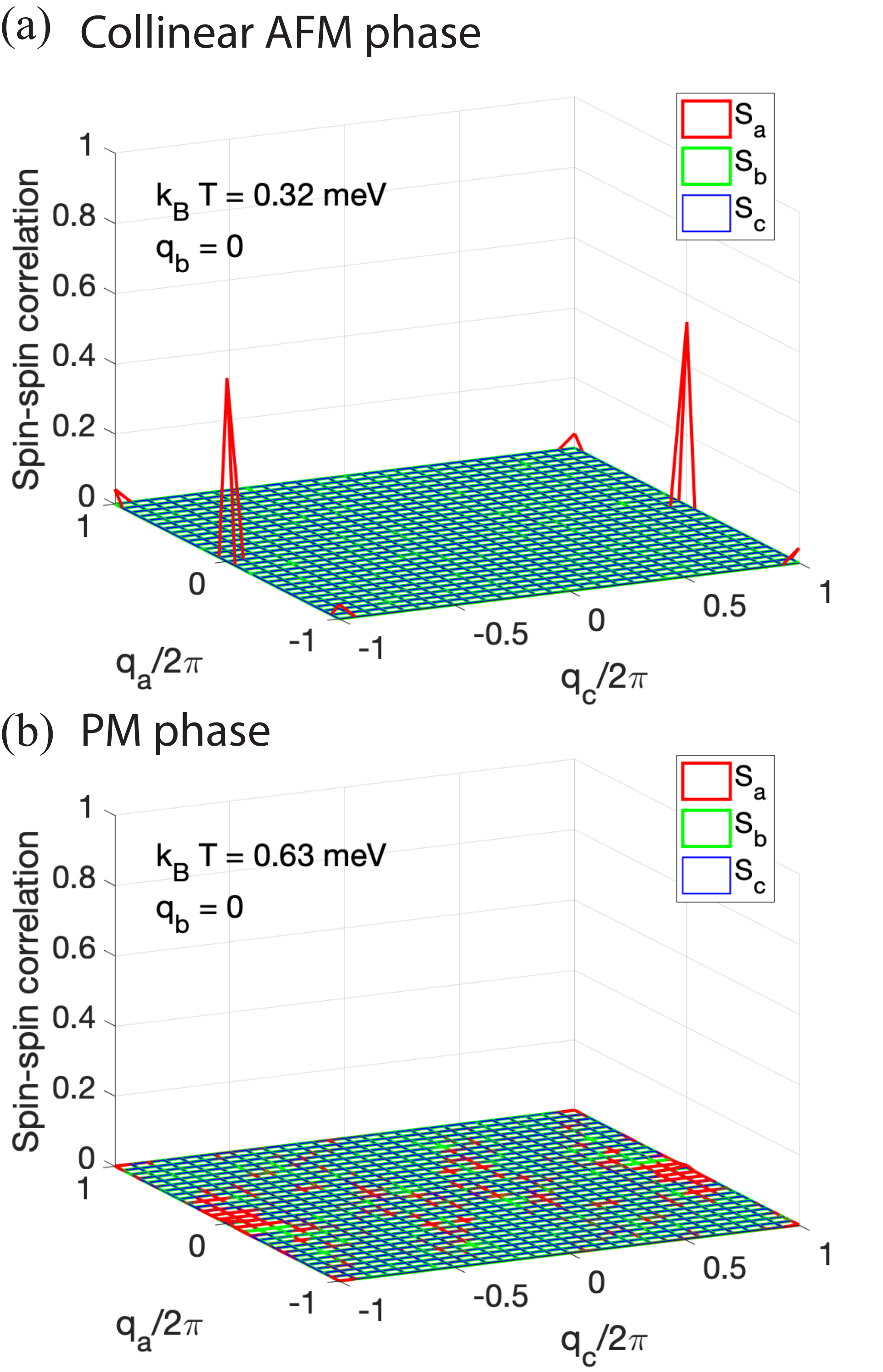}
	\end{center}
	\par
	\caption{\label{structure_exchange4} Calculated spin-spin correlation for collinear AFM phase for (a) $q_{b} = 0$ at $k_{B}T = 0.32$ meV, and PM phase for (b) $q_{b} = 0$ at $k_{B}T = 0.63$ meV.
}
\end{figure}

In \autoref{structure_exchange3}, we present the calculated specific heat, staggered moment of four spins, and spin-helicity. \autoref{structure_exchange4} shows the spin-spin correlation functions at AFM state and PM state. These calculations are performed for $J_{4}/J_{2}$ = 0.74 where the single magnetic phase transition is observed, which serves as as representative result for the cases of $J_{4}/J_{2} < $ 0.95.
\autoref{structure_exchange3} (a) shows the specific heat as a function of temperature at $J_{4}/J_{2}$ = 0.74. A sharp peak around $k_{B}T = 0.60$ meV is observed, signaling a magnetic phase transition below which no more anomalies exist. The calculated spin helicity defined in Eq. (\ref{helicity}) and the staggered magnetic moment defined in Eq. (\ref{staggered}) are shown in \autoref{structure_exchange3} (b) and (c), respectively. Although all components of the spin helicity remains around zero throughout the entire temperature range, we only present $b$-component of the spin helicities whose spin rotational plane is $ac$-plane to compare the case of noncollinear ground state found when $J_{4}/J_{2} >$  0.95. In contrast, the $a$ component of staggered moment begins to grow at the temperature that coincides with the peak position in the specific heat. It approaches 1 with decreasing temperature as the spins fully align along $a$ axis in the staggered manner. 

The spin-spin correlation calculated for $J_{4}/J_{2}$ = 0.74 is shown in \autoref{structure_exchange4}. The spin-spin correlation of the collinear AFM phase at $k_{B}T = 0.32$ meV in \autoref{structure_exchange4} (a) contains the peak of $\alpha = a$ at $q_{c} = 2\pi\left(2n-1\right)$, which also establishes the collinear AFM state with the major spin component along $a$-axis. From the wavevector, all four spins align antiferromagnetically within the unit cell denoted in \autoref{structure_exchange2} and the magnetic and nuclear unit cell are identical. The calculated spin structure agrees well with experimental results as observed for $A =$ K and Rb \cite{K_Rb_spinstructure}. \autoref{structure_exchange4} (b) is the spin-spin correlation in PM phase at $k_{B}T = 0.63$ meV, which shows no correlation among spins as expected for the PM state. The fact that the spin-spin correlation shows the peak with $q_{b} = 0$ indicates the same magnetic structure repeats along $b$-axis. All of our calculations are consistent with the fact that the single magnetic phase transition corresponds to a transition from PM at high temperatures to collinear AFM with spin aligned along $a$-axis at low temperatures. 

\begin{figure}[tbp]
	\linespread{1}
	\par
	\begin{center}
		\includegraphics[width=0.45\textwidth]{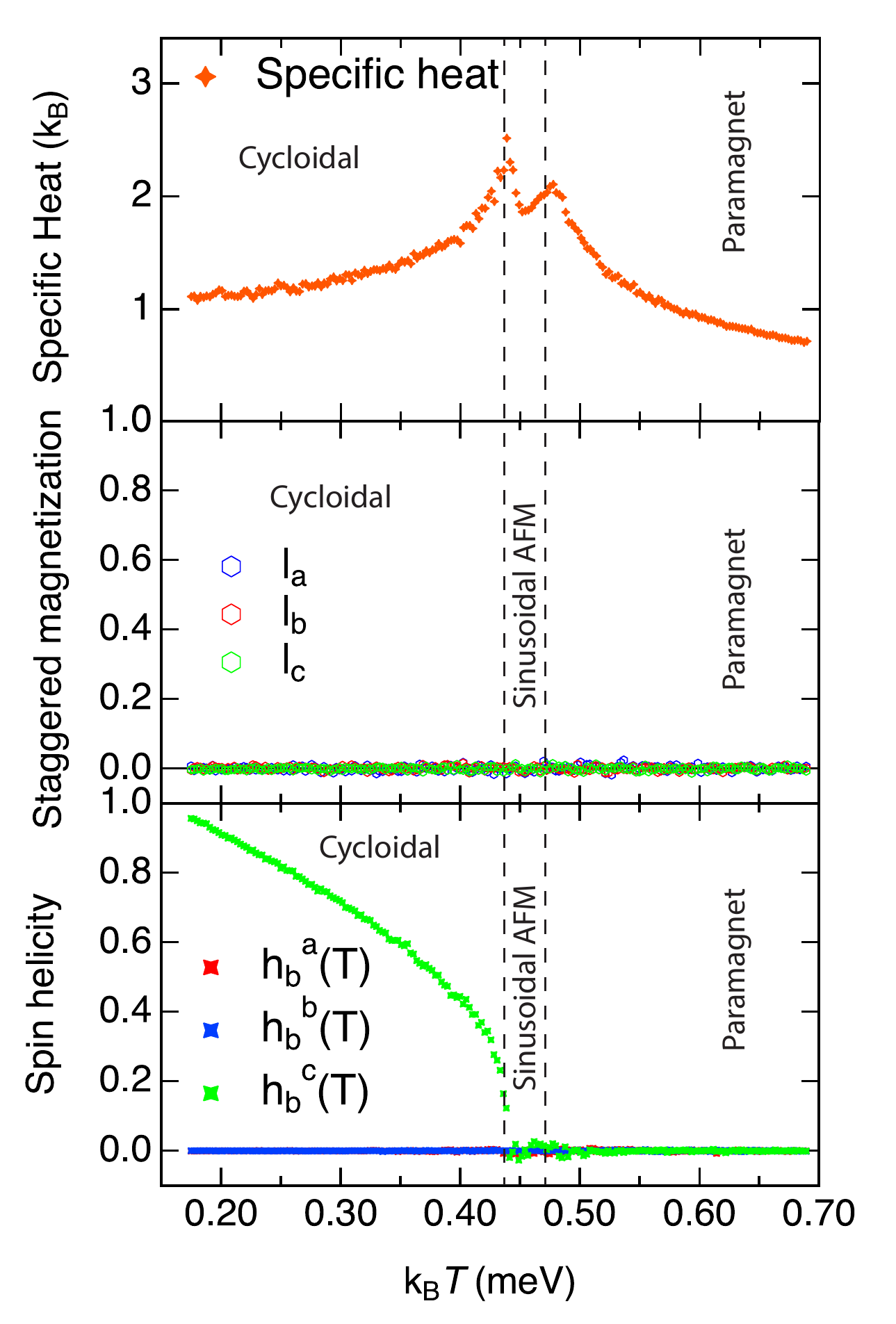}
	\end{center}
	\par
	\caption{\label{structure_exchange5} (color online) Thermodynamic properties of paramagnetic - sinusoidal AFM - cycloidal phase transitions ($J_{4}/J_{2} > $  0.9). Calculated (a) specific heat (\autoref{specificheat}) and (b) staggered magnetic moment (\autoref{staggered}) and spin helicity (\autoref{helicity}) as a function of temperature at zero field at $J_{4}/J_{2}$ = 1.07.}
\end{figure}
As $J_{4}/J_{2}$ increases to 1.07, the degree of frustration increases. The specific heat shows two sharp peaks at $k_{B}T = 0.47$ meV and $k_{B}T = 0.44$ meV as shown in \autoref{structure_exchange5} (a), indicating the two-step phase transition. The calculated staggered moment that serves as the order parameter of collinear AFM phase stays near zero for the entire temperature range as shown in \autoref{structure_exchange5} (b). On the other hand, the spin helicity in \autoref{structure_exchange5} (c) remains near zero in the PM state but begins to rise below $k_{B}T = 0.44$ meV. The fact that the $b$-component of the spin helicity with the propagation vector along $c$ is nonzero suggests that the spin structure of the ground state is cycloidal, rotating in the $ac$-plane and propagating along $c$-axis. This spin state has been experimentally confirmed from the neutron diffraction measurement for the NH$_{4}$ compound \cite{mechanismchange1}. In addition, the electric polarization induced by the inverse DM mechanism points along the $a$ ($\propto {\hat b} \times {\hat c}$)  direction, which has also been confirmed by thermodynamic measurements \cite{Ruby_79K_2}. Thus our calculations also verifies the inverse DM interaction as the microscopic mechanism for the multiferroicity in NH$_{4}$ compound at zero field. The sinusoidal AFM phase is not captured by the staggered moment and spin helicity because it neither has a well-defined staggered moment nor forms a noncollinear phase. 

\begin{figure}[tbp]
	\linespread{1}
	\par
	\begin{center}
		\includegraphics[width=0.45\textwidth]{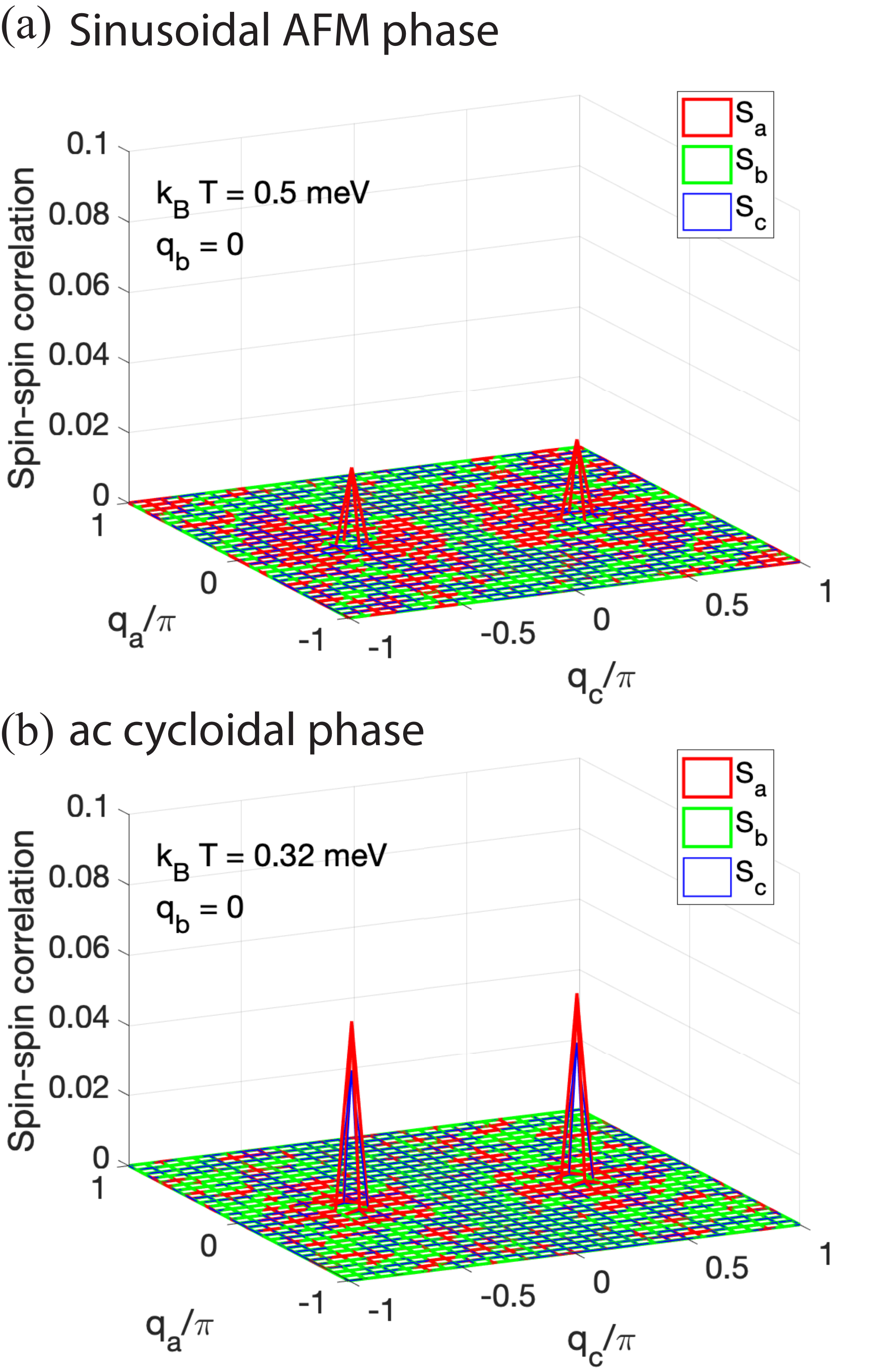}
	\end{center}
	\par
	\caption{\label{structure_exchange6} (color online) Calculated spin-spin correlation for sinusoidal AFM phase for (a) $q_{b} = 0$ (b) at $k_{B}T = 0.5$ meV, and ac cycloidal phase for (b) $q_{b} = 0$ at $k_{B}T = 0.32$ meV.
}
\end{figure}
\autoref{structure_exchange6} shows the spin-spin correlation calculated at $J_{4}/J_{2}$ = 1.07, where the geometrical frustration is significant. \autoref{structure_exchange6} (a) is the spin-spin correlation as a function of $q_{a}$ and $q_{c}$ with $q_{b} = 0$ at $k_{B}T = 0.45$ meV, which is the intermediate phase between paramagnetic phase and $ac$ cycloidal phase. In this phase, the spin-spin correlation function exhibits a sharp peak only in $S_{a}$ component around ${\bf q} \sim \left[0, 0, 0.25\right]$. This suggests that the intermediate phase is a sinusoidal AFM phase in which the $b$ and $c$ components of spins are not correlated but $a$ components are correlated and modulating along $c$ direction with $q_{c} \sim 0.25$. Thus, roughly four spins form one period of the sinusoidal AFM state along $a$-axis. The sinusoidally-modulated spin structure repeats along the $b$ axis ($q_{b} = 0$). Experimentally, the wavevector is incommensurate $q_{c} \sim 0.23$ \cite{Ruby_spin_lattice}, which is slightly smaller than what we obtained here. To resolve the discrepancy of the wavevector between our calculation and experimental data, the larger lattice size needs to be simulated. \autoref{structure_exchange6} (b) shows the spin-spin correlation in the $q_{a}$ and $q_{c}$ plane at $q_{b} = 0$ at $k_{B}T = 0.32$ meV, where the thermodynamic data indicate the $ac$ spin-cycloidal phase. In this phase, the spin-spin correlation calculation exhibits peaks in the $S_{a}$ and $S_{c}$ components around ${\bf q} \sim \left[0, 0, 0.25\right]$. The peak height difference between $S_{a}$ and $S_{c}$ comes from the anisotropy between $a$ and $c$-axis. Therefore the $a$ and $c$ component are correlated in this phase and propagate along $c$ axis, consistent not only with the $ac$-cycloidal phase calculated in our spin-helicity calculation in \autoref{structure_exchange5} (b), but also with the spin structure resolved from neutron diffraction experiments \cite{Ruby_79K_1,Ruby_spin_lattice}.

\section{DISCUSSION}
When $A$ is an alkali metal such as K and Rb, the hydrogen bonds are absent, which results in $J_{4}$ being suppressed compared to $J_{2}$. Thus, the magnetic frustration is reduced although there are still multiple competing isotropic exchange interactions. The ground state then is antiferromagnetic and the direction of the spin component is solely determined by the single-ion anisotropy, which favors the $a$ direction. We believe that any anisotropies induced by DM interactions from the bonds longer than nearest neighbors are too small to explain the anisotropy. On the other hand, when $A$ is NH$_{4}$, hydrogen bonds are formed, which increases $J_{4}$ relative to $J_{2}$ \cite{amanda1}. The resulting triangular geometry via comparable antiferromagnetic exchange interactions of $J_{4}$ and $J_{2}$ promotes a strong geometrical frustration and the noncollinear spin structure is stabilized. In the limiting case when there exist only $J_{2}$ and $J_{4}$, the ground state of the frustrated triangular lattice with Heisenberg spins is the spin cycloidal phase with propagation vector $k = 2\cos^{-1} \left(\frac{J_{2}}{2J_{4}}\right)$. We believe that the other exchange interactions determine the phase difference of the spin cycloids between triangular layers. For instance, the strongest $J_{1}$ exchange interaction connecting two triangular lattice planes tends to make the phase difference between spin cycloidals on different planes $\sim \pi$ \cite{xiaojian_1}.

In (NH$_{4}$)$_{2}$Fe$X_{5}\cdot$H$_{2}$O, the spin cycloid is confined to the $ac$ plane because the $b$ axis is the hard axis. The presence of the odd harmonics in elastic neutron diffraction measurements clearly demonstrates the difference magnetic anisotropy between $a$ and $c$ axes, although inelastic neutron scattering measurement cannot distinguish the difference between them because of the very small energy gap induced by the difference. In our calculations, we also demonstrate that it is required to have different magnetic anisotropy along $a$ and $c$-axis to reproduce the sinusoidal AFM intermediate phase between cycloidal and paramagnetic phase in (NH$_{4}$)$_{2}$Fe$X_{5}\cdot$H$_{2}$O. To resolve the magnetic anisotropy between $a$ and $c$-axes, more sensitive experimental tools are necessary such as EPR and THz spectroscopy.

Based on our result, we can also deduce that the $\sim 30 \%$ difference in $T_{N}$ of Rb$_{2}$FeCl$_{5}\cdot$H$_{2}$O ($T_{N}\sim$ 10 K) and K$_{2}$FeCl$_{5}\cdot$H$_{2}$O ($T_{N}\sim$ 14 K) despite their small lattice parameters difference less than $0.7\%$ is because Rb$_{2}$FeCl$_{5}\cdot$H$_{2}$O has slightly higher $J_{4}/J_{2}$ values, indicative of higher frustration \cite{alkali_neutron1,alkali6}. Cs erythrosiderite has an even lower $T_{N}$. However, the lattice parameters of the Cs compound deviate significantly from those of K and Rb. Thus, the lower $T_N$ could simply result from smaller exchange interactions. Interestingly, our simulation indicates that (NH$_{4}$)$_{2}$FeCl$_{5}\cdot$H$_{2}$O compound exists close to the phase boundary where multiple phases converges as shown \autoref{PD1} (a). Quantum phase transitions should be observable between collinear AFM and noncollinear AFM phases induced by external stimuli such as pressure and magnetic fields, which can be an interesting future study.

Detailed investigations of the effect of magnetic field on NH$_{4}$ compounds have been recently performed. The magnetic field versus temperature phase diagram includes a series of magnetic phase transitions accompanying changes in ferroelectric properties. In particular, multiple transition just below the magnetic saturation is highly interesting but the origin remains elusive \cite{amanda1}. An entanglement of coexisting incommensurate and commensurate phase and their interplay under applied field relevant with lattice strain effect \cite{Weitian1} and subtle changes in the direction of the electric polarization suggest that spin and lattice coupling plays an important role in these erythrosiderites. Our Heisenberg Hamiltonian used in our simulation, which consists of five isotropic exchange interactions and three single anisotropic terms is too simple to explain the effect of spin-lattice constant and the entangled states. More sophisticated model including spin-lattice coupling will be helpful to uncover the fundamental nature of phase transitions under fields. 

\section{CONCLUSION}
In this study, we performed classical Monte Carlo simulations to understand the mechanism to stabilize ground states and classical phase transitions in erythrosiderites. Our simulation results shows that the ground state of $A_{2}$Fe$X_{5}\cdot$H$_{2}$O, where $A$ = Rb, K, and (NH$_{4}$) and $X$ = Cl and Br can be explained by varying the ratio of $J_{4}/J_{2}$ and thus the degree to which the in-plane lattice forms frustrated triangles. In addition, we find that the single-ion anisotropy along $a$ and $c$ must vary in order to give rise to the intermediate-temperature phase in (NH$_{4}$)$_{2}$Fe$X_{5}\cdot$H$_{2}$O compounds. Our study indicates that exotic quantum phase transitions induced by pressure and/or magnetic field are possible in (NH$_{4}$)$_{2}$Fe$X_{5}\cdot$H$_{2}$O and Rb$_{2}$Fe$X_{5}\cdot$H$_{2}$O compounds.
\begin{acknowledgments}
M. L acknowledges V. Zapf, J. Musfeldt, R. Fishman, X. Bao, and W. Tian for valuable discussion, Óscar Fabelo for providing the detailed crystal structure at 2 K, and S. Bae and C. Yaya for their advice for figures. This work was supported by the Center for Molecular Magnetic Quantum Materials, an Energy Frontier Research Center funded by the U.S. Department of Energy, Office of Science, Basic Energy Sciences under Award No. DE-SC0019330.
\end{acknowledgments}

\end{document}